\documentclass[12pt]{article}%
\usepackage{ amsmath, graphics, rotating}%

\begin{document}
\large
\begin{center}{\large\bf WHY IS QUANTUM PHYSICS BASED 
ON COMPLEX NUMBERS?}
\end{center}
\vskip 1em \begin{center} {\large Felix M. Lev} \end{center}
\vskip 1em \begin{center} {\it Artwork Conversion Software Inc.,
1201 Morningside Drive, Manhattan Beach, CA 90266, USA 
(E-mail:  felixlev@hotmail.com)} \end{center}
\vskip 1em

{\it Abstract:}
\vskip 0.5em

The modern quantum theory is based on the assumption
that quantum states are represented by elements of a
complex Hilbert space. It is expected that in future
quantum theory the number field will be not postulated 
but derived from more general principles. We consider 
the choice of the number field in a quantum theory 
based on a finite field. We assume that the symmetry 
algebra is 
the finite field analog of the de Sitter algebra 
so(1,4) and consider spinless
irreducible representations of this algebra.
It is shown that the finite field analog of complex 
numbers is the minimal extension of the residue field 
modulo $p$ for which the representations are
fully decomposable. 

\section{The statement of the problem}
\label{S1} 
 
The modern quantum theory is based on the 
following assumptions:

\begin{itemize}

\item {\it Assumption 1: Quantum states are represented 
by elements of a (projective) complex Hilbert space.}

\item {\it Assumption 2: Observable physical quantities 
are represented by selfadjoint operators in this space.}

\end{itemize}

The field of complex numbers
is algebraically closed, i.e. any equation of the
$n$th power in this field has precisely $n$ solutions.
As a consequence, any linear operator in a finite
dimensional space has at least one eigenvalue.
However, this is not necessarily the case if the
space is infinite-dimensional.  

The usual motivation of Assumption 2 is that 
since any physical quantity can take only real
values, the spectrum of the corresponding
operator should necessarily be real. According
to the spectral theorem for selfadjoint 
operators in Hilbert spaces, this is indeed the
case. However, detailed arguments given in 
Ref. \cite{JMLL} and references therein show 
that the real spectrum
and Assumption 2 are not necessary for 
constructing meaningful quantum theory.
Note that (by definition) any complex 
number is simply a pair of real numbers, and even
for this reason it is not clear why the case of 
complex spectrum should be excluded. For example, 
the complex spectrum can represent a pair of real
physical quantities. 

In quantum theory it is also postulated that 
the following requirement should be valid:

\begin{itemize}

\item {\it Requirement 1: Any linear operator 
representing a physical quantity should have a 
spectral decomposition.}

\end{itemize}

This implies that one can construct
a basis such that any its element is the eigenvector
of the given operator. As it is usual in quantum 
physics, in the general case the 
basis is understood in the sense of distributions, 
i.e. points belonging to the continuous spectrum 
are also treated as eigenvalues. 

As follows from the spectral theorem, if one
accepts Assumption 2 then Requirement 1 is
satisfied automatically. However, the spectral
decomposition exists not only for selfadjoint
operators; for example, it also exists for unitary
operators. It is also clear that the spectral theorem 
for selfadjoint operators is valid not only in complex 
Hilbert spaces but in real Hilbert spaces as well. 
Therefore the only  motivation of 
Assumption 1 is that quantum theory based on complex 
numbers successfully describes a wide range of physical 
phenomena.

It is reasonable to believe that in future
quantum physics the choice of the number 
field (or body) will be substantiated instead 
of saying that a particular
number field should be chosen because the corresponding
version of quantum theory is in agreement with
experimental data. 

In the literature several possibilities have been
considered when the principle number field is not 
the field of complex numbers. There exists a wide
literature devoted to quaternionic, $p$-adic and
adelic versions of quantum theory. In each case 
the theory has its own interesting
properties but the problem of the motivation
of the choice of the principal number field
remains. There also exists a number of works in which
implications of finite fields in quantum physics are
considered (see e.g. Ref. \cite{Galois}). However, to the best of 
our knowledge, only in Refs. \cite{lev1,lev3} and our 
subsequent publications it has been considered a case when
a finite field is the principle number field in
quantum theory. 

There are several arguments for choosing a finite field as
a principal number field in quantum theory is as
follows. For example, if one accepts that the ultimate quantum
theory should not contain actual infinity at all,
then the only possible choice of a number field is
the choice of a finite field. It is well known  
(see e.g. the standard textbooks \cite{VDW}) 
that any finite field contains $p^n$ elements, where 
$p$ is a prime number and $n$ is a natural number. 
Moreover, the choice of $p$ and $n$ 
defines the finite field uniquely up to 
isomorphism. The number $p$ is called the
characteristic of the finite field.

We use $GF(p^n)$ to denote a finite field containing
$p^n$ elements. As it has been shown in Refs. 
\cite{lev1,lev3} and our subsequent publications
(see e.g. Ref. \cite{tmf}),
a quantum theory based on a Galois field $GF(p^2)$ 
(GFQT) is a natural generalization of the standard 
quantum theory based on complex numbers.

Suppose that in our world the principal number
field in quantum theory is a finite field 
characterized by some value of $p$. Then we still 
have to answer the question whether there exist
deep reasons for choosing this particular
value of $p$ or this is simply an accident
that our Universe has been created with this
value. In any case, if we accept that $p$ is
a universal constant then the problem arises
what the value of $n$ is. In view of the above 
discussion it is desirable not to postulate
that $n=2$ but to find a motivation for such a
choice. By analogy with Assumption 1, we 
accept that 

\begin{itemize}

\item {\it Assumption G1: Quantum states in GFQT 
are represented by elements of a 
linear projective space over a field $GF({p^n})$ 
and physical quantities are represented by 
linear operators in that space.}

\end{itemize}
Then we do not require any analog of Assumption 2.
Instead we accept Requirement 1 which in the case
of the GFQT has the same formulation. In the case
of finite-dimensional spaces, the existence of
the spectral decomposition for some operator $A$ 
means precisely that one can construct a basis 
in the usual sense such that all its elements are 
the eigenvectors of $A$. 

Note that any finite field is not algebraically
closed, i.e. any equation of the $n$th power in
the field does necessarily have $n$ solutions. Moreover,
it can have no solutions at all. For this
reason there is no guarantee that any linear
operator in a space over a finite field has
even one eigenvalue, to say nothing about the
possibility that it has a spectral decomposition.

In the present paper we assume that the symmetry 
algebra in GFQT is the finite
field analog of the de Sitter algebra so(1,4)
and consider spinless irreducible representations
(IRs) of this algebra. {\it The main result of the paper
is the proof that if $p=3\,\,(mod\,\,4)$ then the 
minimal extension of $GF(p)$ for which there exist 10
linearly independent representation operators
satisfying Requirement 1, is the field $GF({p^2})$}. 

The paper is organized as follows. In Sect. \ref{S2}
we describe well known facts about modular 
representations. We also consider
modular IRs of the su(2) algebra and argue that the case
$p=3\,\,(mod\,\,4)$ is more natural than 
$p=1\,\,(mod\,\,4)$. In Sect. \ref{S3} we describe
spinless modular IRs of the so(1,4) algebra and
in Sect. \ref{S4} construct a basis convenient for
investigating the spectrum of the operator $M_{04}$. 
The main result is proved in Sect. \ref{S5},
and in Sect. \ref{S6} we discuss Hermiticity
conditions when IR is supplied by a scalar product.
Finally, Sect. \ref{S7} is discussion.  

\section{Modular representations of Lie algebras}
\label{S2}

In standard quantum theory a Lie algebra over the
field of real numbers is called the symmetry algebra
if the given system is described by
a selfadjoint representation of the algebra in a 
Hilbert space. For this reason we accept the following

{\it Definition: a representation of the Lie algebra
over $GF(p)$ in a space over a finite field is called
fully decomposable if the space of representation
operators has a basis such that all
its elements have a spectral decomposition.}

Then a Lie algebra over $GF(p)$ is the symmetry 
algebra if the system is described by a fully
decomposable representation of the algebra
in a linear space over a finite field.

As noted in Ref. \cite{lev3,tmf},
the correspondence between GFQT and standard
quantum theory exists if quantum states in GFQT
are represented by elements of a space over $GF({p^2})$
and $p$ is very large. The field $GF({p^2})$ is a
quadratic extension of the field $GF(p)$. 

It is well known that the element $p-1\in GF(p)$, 
which can be written simply as -1, is a quadratic 
residue 
if $p=1\,\,(mod\,\,4)$ and quadratic nonresidue
if $p=3\,\,(mod\,\,4)$. Therefore in the latter case
the field $GF({p^2})$ can be treated as a complex
extension of $GF(p)$ (i.e. the elements of $GF({p^2})$
can be formally written as $a+bi$ where $a,b\in GF(p)$
and $i$ formally satisfies the condition $i^2=-1$).
The field $GF(p^2)$ has only one nontrivial
automorphism which will be denoted as 
$z\rightarrow {\bar z}$
if $z\in GF({p^2})$. In the case $p=3\,\,(mod\,\,4)$, 
it coincides with the standard complex conjugation 
$a+bi\rightarrow a-bi$. 

By analogy with the conventional quantum theory, one
could require that linear spaces V over $GF({p^2})$, 
used for describing physical states in the GFQT, are 
supplied by a scalar product (...,...) such that for 
any $x,y\in V$ and $a\in GF({p^2})$, $(x,y)$ is an 
element of $GF({p^2})$
and the following properties are satisfied: 
\begin{equation}
(x,y) =\overline{(y,x)},\quad (ax,y)=\bar{a}(x,y),\quad 
(x,ay)=a(x,y)
\label{1}
\end{equation}  

In the general case a scalar product in $V$ does not 
define any positive definite metric, and hence a
probabilistic interpretation exists only for a
subset in $V$ \cite{lev3}. In particular, $(e,e)=0$ 
does not necessarily imply that $e=0$. The quantity
$(e,e)$ can be called the norm (or norm squared) of
the element $e$, but in $GF(p)$ the separation of 
elements into positive and negative does not have the 
same meaning as in the usual case. 

If $A_1$ and $A_2$ are linear operators in $V$ such that
\begin{equation}
(A_1x,y)=(x,A_2y)\quad \forall x,y\in V
\label{2}
\end{equation}
they are said to be adjoint: $A_2=A_1^*$. Then
$A_1^{**}=A_1$ and $A_2^*=A_1$. If $A=A^*$ then, 
by analogy with the standard cases, we can say that
operator $A$ is Hermitian. 
If $Ae=ae$, $a\in GF({p^2})$ and $e\neq 0$ then the element 
$e$ is called the eigenvector of the operator $A$ with the 
eigenvalue $a$. If $A^*=A$ then by analogy with the 
standard case $a\in GF(p)$ if $(e,e)\neq 0$. However, if
$(e,e)=0$ then such a conclusion cannot be drawn. 

If $A$ is a Hermitian operator such that
\begin{eqnarray}
&&Ae_1=a_1e_1,\quad Ae_2=a_2e_2,\quad (e_1,e_1)\neq 0,\nonumber\\
&& (e_2,e_2)\neq 0, \quad a_1\neq a_2
\label{3}
\end{eqnarray}
then as in the usual case, $(e_1,e_2)=0$. 
We will see below that there also exists a possibility
that
\begin{eqnarray}
(e_1,e_1)=(e_2,e_2)=0\quad (e_1,e_2)\neq 0
\label{4}
\end{eqnarray}
In that case it easy to see from Eqs. (\ref{1}) and
(\ref{2}) that ${\bar a}_1=a_2$.

Suppose that the elements $e_1,...e_N$ form a basis
in a space over a finite field and there exists
a scalar product such that $(e_k,e_l)=0$ if $k\neq l$
and $(e_k,e_k)\neq 0$ $\forall k,l=1,2...N$. Then
if $x=c_1e_1+...c_Ne_N$, the coefficient $c_k$ can
be found as $(e_k,x)/(e_k,e_k)$.  

Representations in spaces over a field of nonzero
characteristics are called modular representations.
There exists a wide literature devoted to such 
representations (see e.g. Ref. \cite{Jantzen} and
references therein). In particular, it has
been shown by Zassenhaus \cite{Zass} that all modular 
IRs are finite-dimensional and in numerous papers the 
maximum dimension of such representations is considered.

It is worth noting that usually 
mathematicians consider only representations
over an algebraically closed (infinite) field while 
our approach is different. We consider only finite
fields and investigate what is the minimal
extension of $GF(p)$ such that modular IRs of the
symmetry algebra are fully decomposable. 

Consider, for example, a modular analog of IRs of the
su(2) algebra. Let ${\bf J}=(J_1, J_2, J_3)$ be the 
operator of ordinary rotations in 
the standard theory. If $\hbar /2$ rather than $\hbar$ 
is taken as a unit of measurement of angular momentum 
then the commutation relations for the
components of ${\bf J}$ have the form
\begin{equation}
[J_1,J_2]=2iJ_3,\quad [J_3,J_1]=2iJ_2,\quad 
[J_2,J_3]=2iJ_1
\label{5}
\end{equation}
Define the operators $J_{\pm}$ such that
\begin{equation}
J_1 = J_++J_- \quad J_2 = -i(J_+-J_-)
\label{6}
\end{equation}
Then Eq. (\ref{5}) can be rewritten as
\begin{equation}
[J_3,J_+]=2J_+\quad [J_3,J_-]=-2J_-\quad [J_+,J_-]=J_3
\label{7}
\end{equation}
Since Eq. (\ref{7}) does not contain the quantity $i$,
we now can require that in the modular case
the operators $(J_+J_-J_3)$ act in a space over a
finite field and satisfy the same relations. 

As follows from Eq. (\ref{7}), the operator 
\begin{equation}
K=J_3^2-2J_3+4J_+J_-=J_3^2+2J_3+4J_-J_+
\label{8}
\end{equation}
is the Casimir operator for algebra $(J_+J_-J_3)$. 
Consider a representations containing a vector 
$e_0$ such that
\begin{equation}
J_+e_0=0,\quad J_3e_0=se_0
\label{9}
\end{equation}
where $s\in GF(p)$. Then, as follows from
Eq. (\ref{8}), $e_0$ is the eigenvector of the operator
$K$ with the eigenvalue $s(s+2)$. Denote 
\begin{equation}
e_n=(J_-)^ne_0,\quad n=0,1,2...
\label{10}
\end{equation}
Since $K$ is the Casimir operator, all the $e_n$ are its
eigenvectors with the same eigenvalue $s(s+2)$, and, as 
follows from Eq. (\ref{7}), $J_3e_n=(s-2n)e_n$. 
Hence it follows from Eq. (\ref{8}) that
\begin{equation}
J_+J_-e_n=(n+1)(s-n)e_n
\label{11}
\end{equation}

The maximum value of $n$, $n_{max}$ is defined by
the condition that $J_-e_n=0$ if $n=n_{max}$. This
condition should be compatible with Eq. (\ref{11})
and therefore $n_{max}=s$. It is easy to see that
the elements $e_n$ for $n=0,1,...s$ form a basis
of modular IR and therefore the dimension of 
modular IR with a given $s$ is equal to $s+1$, as
in the standard case. The only difference is that 
in the ordinary case $s$ can be any natural number 
while in the modular case $s$ can take only the values 
of $0,1,...p-1$.

In the standard case the operator $J_3$ is Hermitian
while $J_+^*=J_-$. One can assume that the modular IR
is considered in a space over $GF({p^2})$ and the same
Hermiticity conditions are satisfied. Then it follows 
from Eq. (\ref{11})
that \begin{equation}
(e_{n+1},e_{n+1})=(n+1)(s-n)(e_n,e_n)
\label{12}
\end{equation}
while the elements $e_n$ with the different values
of $n$ are orthogonal to each other.
Therefore, if $(e_0,e_0)\neq 0$ then all the basis 
elements have the nonzero
norm and are orthogonal to each other. However, 
we will not assume in advance that 
modular IRs are considered in a space over 
$GF({p^2})$. As explained above, our goal is to
investigate what is the minimal extension of $GF(p)$
such that modular IRs of the su(2) algebra have three
linearly independent observable operators.

The operator $J_3$ has the spectral decomposition
by construction. Consider now the operator 
$J_1=J_++J_-$ which in the standard theory is the
$x$ component of the angular momentum. We use the
Pochhammer symbol $(a)_l$ to denote 
$a(a+1)\cdots (a+l-1)$ and the standard notation
\begin{equation}
F(a,b;c;z)=\sum_l \frac{(a)_l(b)_lz^l}{l!(c)_l}
\label{13}
\end{equation}
for the hypergeometric series. Let us define
\begin{equation}
e_j^{(x)}=\sum_{k=0}^s \frac{1}{k!}F(-j,-k;-s;2)e_k
\label{14}
\end{equation}
As follows from Eqs. (\ref{10}) and (\ref{11}),
\begin{eqnarray}
&&J_1e_j^{(x)}=\sum_{k=0}^s \frac{1}{k!}
\{F(-j,-k-1;-s;2)(s-k)+\nonumber\\
&&F(-j,-k+1;-s;2)k\}e_k
\label{15}
\end{eqnarray}
Now we use the following relation between the
hypergeometric functions (see e.g. Ref. \cite{BE}):
\begin{eqnarray}
&&[c-2a-(b-a)z]F(a,b;c;z)+a(1-z)F(a+1,b;c;z)-\nonumber\\
&&(c-a)F(a-1,b,c;z)=0
\label{16}
\end{eqnarray}
Then it follows from Eq. (\ref{15}) that 
$J_1e_j^{(x)}=(s-2j)e_j^{(x)}$. 

A possible way to prove that the elements $e_j^{(x)}$
$(j=0,1,..s)$ form a basis is to find a transformation
inverse to Eq. (\ref{14}), i.e. to express the
elements $e_k$ $(k=0,1,...s)$ in terms of $e_j^{(x)}$.
Let $C_s^j=s!/[j!(s-j)!]$ be the binomial coefficient. Then
the transformation has the form
\begin{equation}
e_k=\frac{s!}{2^s(s-k)!}\sum_{j=0}^s C_s^j
F(-j,-k;-s;2)e_j^{(x)}
\label{17}
\end{equation}
and the proof is as follows. 
First, as follows from Eq.
(\ref{14}), the r.h.s. of Eq. (\ref{17}) contains
$$\sum_{j=0}^s C_s^j F(-j,-k;-s;2)F(-j,-k;-s;2)$$
We represent this sum as a limit of 
$$\sum_{j=0}^s C_s^j F(-j,-k;-s;2)F(-j,-k;-s;2)x^j$$
when $x\rightarrow 1$ and use the formula \cite{BE}
\begin{eqnarray}
&&\sum_{j=0}^s C_s^j F(-j,-k;-s;2)F(-j,-k';-s;2)x^j=\nonumber\\
&&(1+x)^{s-k-k'}(1-x)^{k+k'}F(-k,-k';-s;-\frac{4x}{(1-x)^2})
\label{18}
\end{eqnarray}
As follows from Eq. (\ref{13}), the series for the
hypergeometric function in Eq. (\ref{18}) has the last
term corresponding to $l=min(k,k')$ and this term is
the most singular when $x\rightarrow 1$. Then it is
clear that if $k\neq k'$, the r.h.s. of Eq. (\ref{18}) is 
equal to zero in the limit $x\rightarrow 1$ while if
$k=k'$ then the limit is equal to $2^s(s-k)!/s!$.
This completes the proof of Eq. (\ref{17}) and
we conclude that the operator $J_1$ has the
spectral decomposition even without extending the
field $GF(p)$. 

Consider now the operator $J_+-J_-$. Since in the
standard theory (see Eq. (\ref{6})) it is equal to
$-iJ_2$ where $J_2$ is the $y$ projection of the
angular momentum, one might expect that in the
modular case $J_+-J_-$ has a spectral
decomposition only if $GF(p)$ is extended.

Consider first the simplest nontrivial case when
$s=1$ ($s=1/2$ in the standard units). Then, as 
follows from Eqs. (\ref{10}) and
(\ref{11}), the characteristic equation for the
operator $J_+-J_-$ is $\lambda^2 = -1$. In the case 
$p=3\,\,(mod\,\, 4)$ this
equation can be solved only by extending $GF(p)$.
However, if $p=1\,\,(mod\,\, 4)$, the equation
has solutions in $GF(p)$ and hence no extension
of $GF(p)$ is needed to ensure the spectral
decomposition of the operator $J_+-J_-$.
Nevertheless, if $p$ is very large and 
$p=1\,\,(mod\,\, 4)$ then 
the quantities $\lambda$ satisfying    
$\lambda^2 = -1=p-1$ in $GF(p)$ are very large 
(at least of order $\sqrt{p}$). This obviously 
contradicts experimental data since in the IR 
of the su(2) algebra with $s=1$ no quantities with
such large eigenvalues have been observed.

We conclude that the case
$p=1\,\,(mod\,\, 4)$ is probably incompatible
with the existing data. For this reason we
will consider only quadratic extensions of
$GF(p)$ in the case $p=3\,\,(mod\,\, 4)$. Then
by analogy with the above discussion one can
prove that the elements
\begin{equation}
e_j^{(y)}=\sum_{k=0}^s \frac{1}{k!}F(-j,-k;-s;2)i^ke_k
\label{19}
\end{equation}
are the eigenvectors of the operator $J_+-J_-$
with the eigenvalues $i(s-2j)$ and they form the
basis in the representation space.

Our final conclusions in this section are as
follows. If quantum theory is based on a finite
field then the number $p$ representing the 
characteristic of the field is such that 
$p=3\,\,(mod\,\, 4)$ rather than 
$p=1\,\,(mod\,\, 4)$. Then the
complex extension of $GF(p)$ guarantees that 
modular IRs of the su(2) algebra are fully 
decomposable. 

\section{Modular IRs of the so(1,4) algebra}
\label{S3}

\begin{sloppypar}
In standard quantum theory one can choose
the units in which  $\hbar/2=c=1$. Then the assumption
that the de Sitter algebra so(1,4) is the symmetry
algebra implies that its  representation operators 
$M^{ab}$ ($a,b=0,1,2,3,4$, $M^{ab}=-M^{ba}$) are 
Hermitian and satisfy the
commutation relations
\begin{equation}
[M^{ab},M^{cd}]=-2i (\eta^{ac}M^{bd}+\eta^{bd}M^{as}-
\eta^{ad}M^{bc}-\eta^{bc}M^{ad})
\label{20}
\end{equation}
where $\eta^{ab}$ is the diagonal metric tensor such that
$\eta^{00}=-\eta^{11}=-\eta^{22}=-\eta^{33}=-\eta^{44}=1$.
\end{sloppypar}

One could define the de Sitter invariance in
GFQT by saying that the operators $M^{ab}$ describing
the system act in a space $V$ over
$GF({p^2})$ and satisfy the same relations (\ref{20}) in
that space. However, as noted in Sect. \ref{S1}, our
goal is not to postulate the choice of the number field
but substantiate it from the requirement that
modular IRs of the symmetry algebra
are fully decomposable. Since su(2) is a subalgebra of
so(1,4), it follows from the results of the preceding
section, one cannot obtain a fully decomposable IR
without extending the field $GF(p)$. However, we do not
know yet, whether the complex extension will be
sufficient. For this reason it is desirable not
to fix the field immediately but assume only that it
is an extension of $GF(p)$. In this case we first have
to investigate what conclusions can be drawn without
assuming the existence of any scalar product and
Hermiticity requirements.

In the standard theory, instead  of $M^{ab}$ one can
work with the set of operators 
$({\bf J}',{\bf J}",R_{jk})$ ($j,k=1,2$).  
Here ${\bf J}'$ and ${\bf J}"$ are two independent 
su(2) algebras (i.e. $[{\bf J}',{\bf J}"]=0$)
described by Eqs. (\ref{6}) and (\ref{7}).
The commutation relations of the operators ${\bf J}'$ and
${\bf J}"$  with the $R_{jk}$ have the form
\begin{eqnarray}
&&[J_3',R_{1j}]=R_{1j},\quad [J_3',R_{2j}]=-R_{2j},\quad
[J_3",R_{j1}]=R_{j1},\nonumber\\
&& [J_3",R_{j2}]=-R_{j2},\quad
[J_+',R_{2j}]=R_{1j},\quad [J_+",R_{j2}]=R_{j1},\nonumber\\
&&[J_-',R_{1j}]=R_{2j},\quad [J_-",R_{i1}]=R_{i2},\quad
[J_+',R_{1j}]=\nonumber\\
&&[J_+",R_{j1}]=[J_-',R_{2j}]=[J_-",R_{j2}]=0,\nonumber\\
\label{21}
\end{eqnarray}
and the commutation relations of the operators 
$R_{jk}$ 
with each other have the form
\begin{eqnarray}
&&[R_{11},R_{12}]=2J_+',\quad 
[R_{11},R_{21}]=2J_+",\nonumber\\
&& [R_{11},R_{22}]=-(J_3'+J_3"),\quad 
[R_{12},R_{21}]=J_3'-J_3"\nonumber\\
&& [R_{11},R_{22}]=-2J_-",\quad [R_{21},R_{22}]=-2J_-'
\label{22}
\end{eqnarray}
Then, if ${\bf M}=\{M^{23},M^{31},M^{12}\}$,
${\bf B}=-\{M^{14},M^{24},M^{34}\}$ and  
the relation between the sets 
$({\bf J}',{\bf J}",R_{jk})$ and
$M^{ab}$  is given by
\begin{eqnarray}
&&{\bf M}={\bf J}'+{\bf J}", \quad {\bf B}={\bf J}'-{\bf J}",
\quad M_{01}=\imath (R_{11}-R_{22}), \nonumber\\
&& M_{02}=R_{11}+R_{22}, \quad  
M_{03}=-i(R_{12}+R_{21}),\nonumber\\
&&  M_{04}=R_{12}-R_{21},
\label{23}
\end{eqnarray}
one can directly verify that Eq. (\ref{20}) 
follows from Eqs. (\ref{7}), (\ref{21}), (\ref{22}),
(\ref{23}) and {\it vice versa}.

Since Eqs. (\ref{7}), (\ref{21}) and (\ref{22}) do not 
contain the quantity $i$, one can define the de Sitter
invariance in the modular case by requiring that
the system is described by the operators
$({\bf J}',{\bf J}",R_{jk})$ 
($j,k=1,2$) satisfying these expressions. 
At first glance, these relations might seem 
rather chaotic, but they are natural in the Weyl basis 
of the so(1,4) algebra.

Proceeding from the method of su(2)$\times$su(2) shift
operators, developed by Hughes \cite{Hug} for constructing
standard unitary IRs of the group SO(5), and following 
Ref. \cite{lev3}, we now give a description of
modular IRs of the so(1,4) algebra. 

Consider the space of maximal  $su(2)\times su(2)$  
vectors, i.e.  such vectors $x$ that $J_+'x=J_+"x=0$. Then 
it follows from Eqs. (\ref{7}), (\ref{21}), (\ref{22}) 
that the operators
\begin{eqnarray}
&&A^{++}=R_{11}\quad  A^{+-}=R_{12}(J_3"+1)-
J_-"R_{11},  \nonumber\\
&&A^{-+}=R_{21}(J_3'+1)-J_-'R_{11}\nonumber\\
&&A^{--}=-R_{22}(J_3'+1)(J_3"+1)+J_-"R_{21}(J_3'+1)+\nonumber\\
&&J_-'R_{12}(J_3"+1)-J_-'J_-"R_{11}
\label{24}
\end{eqnarray}
act invariantly on this space.
The notations are related to the property  that
if $x^{kl}$  ($k,l>0$) is the maximal su(2)$\times$su(2)
vector and simultaneously
the eigenvector of operators $J_3'$ and $J_3"$ with the 
eigenvalues $k$ and $l$, respectively, then  $A^{++}x^{kl}$
is  the  eigenvector  of  the  same
operators with the values $k+1$ and $l+1$, $A^{+-}x^{kl}$ - the
eigenvector  with
the values $k+1$ and $l-1$, $A^{-+}x^{kl}$ - 
the eigenvector with the values  $k-1$ and $l+1$ and 
$A^{--}x^{kl}$ - the eigenvector with the
values $k-1$ and $l-1$.

As follows from the results of the preceding
section, the vector $x_{\alpha\beta}^{kl}
=(J_-')^{\alpha}(J_-")^{\beta}x^{kl}$ is
the eigenvector of the operators $J_3'$ and $J_3"$ with 
the eigenvalues $k-2\alpha$ and $l-2\beta$, respectively. 
Since 
$${\bf J}^2=J_3^2-2J_3+4J_+J_-=J_3^2+2J_3+4J_-J_+$$
is the Casimir operator for the ${\bf J}$  algebra,
it  follows  in  addition that 
\begin{equation}
{\bf J}^{'2}x_{\alpha\beta}^{kl}=k(k+2)x_{\alpha\beta}^{kl},\quad
{\bf J}^{"2}x_{\alpha\beta}^{kl}=l(l+2)x_{\alpha\beta}^{kl}
\label{25}
\end{equation}
\begin{equation}
J_+'x_{\alpha\beta}^{kl}=\alpha (k+1-\alpha)
x_{\alpha -1,\beta}^{kl},
\quad  J_+"x_{\alpha\beta}^{kl}=\beta (l+1-\beta)
x_{\alpha,\beta -1}^{kl}
\label{26}
\end{equation}
From these formulas it follows that the action of 
the operators ${\bf J}'$
and ${\bf J}"$ on $x^{kl}$  generates a space with the  
dimension $(k+1)(l+1)$  and  the
basis $x_{\alpha\beta}^{kl}$ ($\alpha=0,1,...k$, 
$\beta=0,1,...l$).

The Casimir operator of the second order for the 
algebra (\ref{20}) can be written as
\begin{eqnarray}
&&I_2 =-\frac{1}{2}\sum_{ab} M_{ab}M^{ab}=\nonumber\\
&&4(R_{22}R_{11}-R_{21}R_{12}-J_3')-
2({\bf J}^{'2}+{\bf J}^{"2})
\label{27}
\end{eqnarray}
and the basis in the representation space
can be explicitly constructed assuming that there exists a
vector $e^0$ which is the maximal su(2)$\times$su(2)
vector such that
\begin{eqnarray}
&&{\bf J}'e^0=J_+"e^0=0,\quad J_3"e^0 =se^0, \nonumber\\
&&I_2e^0 =[w-s(s+2)+9] e^0
\label{28}
\end{eqnarray}
(see Ref. \cite{lev3} for details)
where $w,s\in GF(p)$. 

Define the vectors
\begin{equation}
e^{nr}=(A^{++})^n(A^{+-})^re^0
\label{29}
\end{equation}
Then a direct calculation using Eqs. (\ref{7}),
(\ref{21}), (\ref{22}), (\ref{24}), (\ref{27}), (\ref{28}), 
and (\ref{29}) gives
\begin{equation}
A^{--}A^{++}e^{nr}=-\frac{1}{4}(n+1)(n+s+2)[w+(2n+s+3)^2]e^{nr}
\label{30}
\end{equation}
\begin{equation}
A^{-+}A^{+-}e^{nr}=-\frac{1}{4}(r+1)(s-r)[w+1+(2r-s)
(2r+2-s)]e^{nr}
\label{31}
\end{equation}
As follows from the last expression, $r$ can take only the 
values of $0,1,....s$, as well as in the standard theory.
At the same time, as follows from Eq. (\ref{30}),
the possible values of $n$ in the
modular case are not the same as in the standard theory.
Indeed, in the standard case 
the possible values of $n$ are obviously $n=0,1,2,...\infty$
and therefore IR is infinite dimensional.
On the contrary, in the modular case $n$ can
take only the values $0,1,...n_{max}$ where the maximum 
value of $n$, $n_{max}$ can be found as follows. By definition
of the operator $A^{++}$, $A^{++}e^{nr}=0$ if $n=n_{max}$.
This relation should not contradict Eq. (\ref{30}) if 
$n=n_{max}$. Therefore $n_{max}$ is the least number
satisfying the congruence modulo $p$
\begin{equation}
(n_{max}+1)(n_{max}+s+2)[w+(2n_{max}+s+3)^2]=0
\label{32}
\end{equation}  
In particular, the IR 
is necessarily finite dimensional in agreement with the
Zassenhaus theorem \cite{Zass}. 

In the standard
theory $w=\mu^2$ where $\mu$ is the particle dS mass
(see e.g. Ref. \cite{jpa}) but in GFQT the element 
$w\in GF(p)$ 
is not necessarily a square in $GF(p)$. Since -1 is a 
quadratic residue in $GF(p)$
if $p=1\,\,(mod\,\,4)$ and quadratic nonresidue if
$p=3\,\,(mod\,\,4)$ then in the latter case 
(which is of only interest for us in view of the
results of the preceding section) the equality 
$w+(2n_{max}+s+3)^2=0$ in $GF(p)$ is 
possible only if $w$ is a quadratic nonresidue.
We will see below that only this condition is
consistent. Then $w=-{\tilde \mu}^2$ 
where for ${\tilde\mu}$ obviously two solutions
are possible. 

Consider for simplicity the case
$s=0$. Then $n_{max}$ should satisfy one
of the conditions $2n_{max}+3=\pm {\tilde\mu}$
and therefore there exist two solutions for
$n_{max}$. We should choose one with
the lesser value of $n_{max}$ and, as follows
from Eq. (\ref{32}), this value should be less
than $p-2$. Let us assume that both, 
${\tilde\mu}$ and $-{\tilde\mu}$ are
represented by $0,1,...(p-1)$. 
Then if ${\tilde\mu}$ is odd, 
$-{\tilde\mu}=p-{\tilde\mu}$ is
even and {\it vice versa}. We choose
the odd number as ${\tilde\mu}$. Then
the two solutions are 
$n_1=({\tilde\mu}-3)/2$ and      
$n_2=p-({\tilde\mu}+3)/2$. It is obvious
that $n_1<n_2$ and $n_1<p-2$. Therefore
\begin{equation}
n_{max}=({\tilde\mu}-3)/2 
\label{33}
\end{equation} 
In particular, this quantity satisfies
the condition $n_{max}\leq (p-5)/2$.

Since $e^{nr}$ is the maximal $su(2)\times su(2)$ 
vector with the
eigenvalues of the operators ${\bf J}'$ and 
${\bf J}"$ equal to
$n+r$ and $n+s-r$, respectively, then
as a basis of the representation space one can take 
the vectors
$e_{\alpha\beta}^{nr}=(J_-')^{\alpha} (J_-")^{\beta}e^{nr}$
where, for the given $n$ and $s$, the quantity $\alpha$ can
take the values of $0,1,...n+r$ and $\beta$ - the values 
of $0,1,...n+s-r$. 

If $s=0$ then there
exist only the maximal $su(2)\times su(2)$ vectors $x^{kl}$
with $k=l$ and therefore the basis of the  representation  
space  is
formed by the vectors $e_{\alpha\beta}^n\equiv 
e_{\alpha\beta}^{n0}$
where $n=0,1,2,...$; $\alpha,\beta=0,1,...n$. The explicit 
expressions for the  action  of the operators $R_{jk}$ in this
basis can be calculated by using Eqs. (\ref{21}), 
(\ref{22}) (\ref{29}-\ref{31}) and the result is
\begin{eqnarray}
&& R_{11}e_{\alpha\beta}^n=\frac{(n+1-\alpha)(n+1-\beta)}
{(n+1)^2}e_{\alpha\beta}^{n+1}+ \nonumber\\
&&\frac{\alpha\beta n}{4(n+1)}
[w+(2n+1)^2]e_{\alpha -1,\beta-1}^{n-1},\nonumber\\
&&R_{12}e_{\alpha\beta}^n=\frac{n+1-\alpha}{(n+1)^2}
e_{\alpha,\beta+1}^{n+1}-\frac{\alpha n}{4(n+1)}[w+(2n+1)^2]
e_{\alpha -1,\beta}^{n-1},\nonumber\\
&& R_{21}e_{\alpha\beta}^n=\frac{n+1-\beta}{(n+1)^2}
e_{\alpha+1,\beta}^{n+1}-\frac{\beta n}{4(n+1)}[w+(2n+1)^2]
e_{\alpha,\beta-1}^{n-1},\nonumber\\
&&R_{22}e_{\alpha\beta}^n=\frac{1}{(n+1)^2}
e_{\alpha+1,\beta+1}^{n+1}+\nonumber\\
&&\frac{n}{4(n+1)}[w+(2n+1)^2]e_{\alpha \beta}^{n-1}
\label{34}
\end{eqnarray}

We can now prove that the only consistent case when
Eq. (\ref{32}) is satisfied is that $w$ is a
quadratic nonresidue and $n_{max}=p-2$ is
impossible. Indeed, if $n_{max}$ is a maximum 
possible value of $n$ then, as
follows from Eq. (\ref{34}), 
$R_{11}e^{n_{max}}=R_{12}e^{n_{max}}=0$.
Since $e^{n_{max}}=e^{n_{max}}_{00}$, it
follows from Eq. (\ref{27}),
$I_2e^{n_{max}}=-4n_{max}(n_{max}+3)e^{n_{max}}$. 
This condition
can be compatible with $I_2e^{n_{max}}=(w+9)e^{n_{max}}$ 
(see Eq. 
(\ref{28})) only if $w+(2n_{max}+3)^2=0$.

As follows from the results of the preceding section,
from ${\bf J}'$ and ${\bf J}"$ one can construct six 
linearly independent operators having a spectral
decomposition if $GF(p)$ is extended to $GF({p^2})$
in the case $p=3\,\,(mod\,\,4)$. Our goal is to
prove that such an extension is sufficient to ensure
a possibility of constructing additional four 
independent operators from the $R_{jk}$ such that
all of them have a spectral decomposition. 

For simplicity we consider only the spinless case.
Then, as follows from Eq. (\ref{28}), the vector
$e_0$ is annihilated by all the representation
operators of the so(4)=su(2)$\times$su(2) algebra.
Therefore all the operators $M_{0\nu}$ ($\nu =1,2,3,4$)
are on equal footing and it is sufficient to prove 
that the operator $M_{04}$ has a spectral decomposition. 

\section{$M_{04}$ operator in the $({\bf J}^2,J_3)$
basis}
\label{S4}

We now use ${\bf J}$ to denote ${\bf J}'+{\bf J}"$.
In the standard theory ${\bf J}$ is the angular
momentum corresponding to conventional 
three-dimensional rotations. In the modular case
the set ${\bf J}$ is defined by the operators
$(J_+J_-J_3)$ satisfying the commutation relations
(\ref{7}) since ${\bf J}'$ and  ${\bf J}"$
satisfy these relations and $[{\bf J}',{\bf J}"]=0$.
Since $M_{04}=R_{12}-R_{21}$ (see Eq. (\ref{23}))
then, as follows from Eq. (\ref{21}), 
$[M_{04},{\bf J}]=0$. Therefore for investigating 
operator $M_{04}$ it is convenient to decompose
the representation space into subspaces such that
all the elements belonging to the same subspace
are the eigenvectors of the operators ${\bf J}^2$
and $J_3$ with the same eigenvalue. 

Since $e^n$ satisfies the conditions $J_+'e^n=J_+"e^n=0$
by construction, it also satisfies $J_+e^n=0$
and $J_3e^n=2ne^n$. Therefore
the elements $(J_-)^le^n$ $(l=0,1,...2n)$ form a
subspace corresponding to IR of the su(2) algebra
with the spin $s=2n$. 

\begin{sloppypar}
Let now $e(n,k)$ be an element satisfying the
conditions $J_+e(n,k)=0$ and $J_3e(n,k)=2(n-k)e(n,k)$. 
Then the elements $e(n,k,l)=(J_-)^le(n,k)$ $(l=0,1,...2(n-k))$ 
form a subspace corresponding to IR of the su(2) algebra
with the spin $s=2(n-k)$.
We define 
\begin{equation}
e(n,k)=B(n,k)e(n,0)
\label{35}
\end{equation}
where $e(n,0)=e^n$ and
\begin{eqnarray}
B(n,k)=\frac{1}{(n!)^2}\sum_{l=0}^k (-1)^{k-l}C_k^l(n-l)!
(n+l-k)!(J_-')^l(J_-")^{k-l}
\label{36}
\end{eqnarray}
Then we have to prove that $J_+e(n,k)=0$ for
$k=1,2,...n$.
A direct calculation using Eqs. (\ref{7}), (\ref{35}) 
and (\ref{36}) gives
\begin{equation}
J_+'e(n,k)=ke(n,k-1)\quad J_+"e(n,k)=-ke(n,k-1)
\label{37}
\end{equation}
Therefore
\begin{equation}
(J_+'+J_+")e(n,k)=0\quad (J_-'-J_+")e(n,k)=2ke(n,k-1)
\label{38}
\end{equation}
\end{sloppypar}

It has been shown in the preceding section, that
acting by $J_-'$ and $J_-"$ on the element $e^n$
one can obtain a subspace with the dimension $(n+1)^2$
and the basis $e^n_{\alpha\beta}$. On the other hand,
as shown in this section, in such a way it is also
possible to obtain a subspace with the basis $e(n,k,l)$
where at fixed $n$ and $k$, $l$ takes the values
$0,1,...2(n-k)$ and at a fixed $n$, $k$ takes the values
$0,1,...n$. The problem arises whether the 
elements $e(n,k,l)$ $(n=0,1,...n_{max})$ also form a 
basis in the representation space.

In the standard theory the proof follows from the fact
that the dimension of the subspace generated by the
elements $e(n,k,l)$ at different values of $k$ and $l$
is also equal to $(n+1)^2$ since
$$\sum_{k=0}^n [2(n-k)+1]=(n+1)^2$$ 
Moreover, as follows from Eq.
(\ref{38}), any element $e(n,k,l)$ can be chosen as
a cyclic element of IR. 
It will also be shown in Sect.
\ref{S6} that it is possible to define a scalar 
product in the representation space such that the both 
basis's, $\{e^n_{\alpha\beta}\}$, and $\{e(n,k,l)\}$
satisfy the following property: their different
elements are orthogonal while the norm of each
element is not equal to zero.

In the modular case the situation might be a bit more
complicated. As shown in Sect. \ref{S2}, the dimension
of modular IR of the su(2) algebra characterized by
the value $J$ is equal to $J+1$ only if $J$ is one
of the values $0,1,...p-1$. Therefore the dimension
of IR characterized by the values of $n$ and $k$ is
$2(n-k)+1$ only if $(n-k)\leq (p-1)/2$. For example,
if $n-k=(p+1)/2$, IR corresponds to $J=1$ and has
the dimension 2. This examples shows that in the
modular case $J$ is not necessarily even in the
spinless case. 

As noted in the preceding section, the
quantity $w$ should be a
quadratic nonresidue and then the value of $n_{max}$
is necessarily less than $(p-3)/2$. In that case
all the possible values of $J$ are even and
the dimension of IR characterized by the
maximal weight $J$ is $J+1$, as well as in the
standard theory. Therefore the subspace generated 
by the elements $e(n,k,l)$ at different values
of $k$ and $l$ 
also has the dimension $(n+1)^2$ and all the
the elements $e(n,k,l)$ form a basis in the
representation space. 

Let $V(j)$ be a subspace generated by the elements
$e(n,k)$ such that the value of $j=J/2=(n-k)$ is the
same for all the basis elements. Then they satisfy 
the conditions $J_+e(n,k)=0$ and
$J_3e(n,k)=2je(n,k)$. The basis in $V(j)$ is formed
by the elements $e(n,n-j)$ $(n=j, j+1,...n_{max})$.
One can also define subspaces $V(j,k)=(J_-)^kV(j)$.
It is clear that the representation space can
be decomposed into the subspaces $V(j,k)$ such that
$k=0,1,...2j$ if $j$ is fixed and $j=0,1,..n_{max}$.
Since $M_{04}$ commutes with ${\bf J}$, each
subspace $V(j,k)$ is invariant under the action of
$M_{04}$ and for a fixed $j$ the spectrum of $M_{04}$
in all the subspaces $V(j,k)$ is the same. Therefore
for investigating operator $M_{04}$ it is 
sufficient to consider its action in subspaces $V(j)$.

Suppose that $j$ is fixed and denote 
$E_n=(-1)^ne(n+j,n)$.
Then the elements $E_n$ $(n=0,1,...n_{max}-j)$ form
a basis in $V(j)$. A direct calculation using Eqs.
(\ref{22}) and (\ref{34}-\ref{36}) shows that the
action of $M_{04}=A$ in $V(j)$ is given by
\begin{equation}
AE_n=E_{n+1}+\frac{n(n+2j+1)}{4(n+j)(n+j+1)}
[w+(2n+2j+1)^2]E_{n-1}
\label{39}
\end{equation}
This expression shows that the matrix of the
operator $A$ has only the following nonzero
elements:
\begin{equation}
A_{n+1,n}=1\quad A_{n,n+1}=\frac{(n+1)(n+2j+2)}{4(n+j+1)(n+j+2)}
[w+(2n+2j+3)^2]
\label{40}
\end{equation}
where $n=0,1,...n_{max}-j$.

Note that the results of this and preceding sections
have been obtained assuming that IR is considered
in a space over a finite field of characteristic $p$
but no concrete choice of the field with
such a characteristic has been made.  
  
\section{Spectrum of the $M_{04}$ operator}
\label{S5}

Consider now the spectrum of the operator having the
matrix (\ref{40}). Let $A(\lambda)$ be the matrix of the
operator $A-\lambda$. We use $\Delta_k^l(\lambda)$
to denote the determinant of the matrix obtained from
$A(\lambda)$ by taking into account only the rows and
columns with the numbers $k,k+1,..l$. Our convention
is that in the matrix $A(\lambda)$ the first row and 
column have the values equal to 0 while the last ones 
have the values equal to $N=n(j)_{max}$ which should
be defined yet.
Therefore the characteristic equation can be written as
\begin{equation}
\Delta_0^N(\lambda)=0
\label{41}
\end{equation}

The matrix $A(\lambda)$ is three-diagonal. It is easy
to see that 
\begin{equation}
\Delta_0^{n+1}(\lambda)=-\lambda \Delta_0^n(\lambda)-
A_{n,n+1}A_{n+1,n}\Delta_0^{n-1}(\lambda)
\label{42}
\end{equation}
Let $\lambda_l$ be a solution of Eq. (\ref{41}). Then
the element
\begin{equation}
\chi(\lambda_l)=\sum_{n=0}^N \{(-1)^n \Delta_0^{n-1}(\lambda_l)E_n/
[\prod_{k=0}^{n-1}A_{k,k+1}]\}
\label{43}
\end{equation}
is the eigenvector of the operator $A$ with the eigenvalue
$\lambda_l$. This can be verified directly by using Eqs.
(\ref{39}-\ref{43}).

To solve Eq. (\ref{41}) we have to find the expressions for
$\Delta_0^n(\lambda)$ when $n=0,1,...N$. It is obvious that
$\Delta_0^0(\lambda)=-\lambda$, and as follows from
Eq. (\ref{40}),
\begin{equation}
\Delta_0^1(\lambda)=\lambda^2-\frac{w+(2j+3)^2}{2(j+2)}
\label{44}
\end{equation}
 
Since $w$ should be a quadratic nonresidue, it can be 
represented as $w=-{\tilde\mu}^2$. Then it can be shown 
that
$\Delta_0^n(\lambda)$ is given by the following 
expressions. If $n$ is odd then
\begin{eqnarray}
&&\Delta_0^n(\lambda)=\sum_{l=0}^{(n+1)/2}C_{(n+1)/2}^l
\prod_{k=1}^l[\lambda^2+({\tilde\mu}-2j-4k+1)^2]
(-1)^{(n+1)/2-l}\nonumber\\
&&\prod_{k=l+1}^{(n+1)/2}\frac{2j+2k+1}{2(j+(n+1)/2+k)}
({\tilde\mu}-2j-4k+1)({\tilde\mu}-2j-4k-1)
\label{45}
\end{eqnarray} 
and if $n$ is even then
\begin{eqnarray}
&&\Delta_0^n(\lambda)=(-\lambda)\sum_{l=0}^{n/2}C_{n/2}^l
\prod_{k=1}^l[\lambda^2+({\tilde\mu}-2j-4k+1)^2]
(-1)^{n/2-l}\nonumber\\
&&\prod_{k=l+1}^{(n+1)/2}\frac{2j+2k+1}{2(j+n/2+k+1)}
({\tilde\mu}-2j-4k-1)({\tilde\mu}-2j-4k-3)
\label{46}
\end{eqnarray}
Indeed, for $n=0$ Eq. (\ref{46}) is compatible with
$\Delta_0^0(\lambda)=-\lambda$, and for $n=1$ Eq. (\ref{45})
is compatible with Eq. (\ref{44}). Then one can
directly verify that Eqs. (\ref{45}) and (\ref{46})
are compatible with Eq. (\ref{42}).

As noted in the preceding section, $N$ should be such that 
$N\leq n_{max}-j$ where 
$w+(2n_{max}+3)^2=0$ in the spinless case. On the
other hand, $N$ is the greatest value of $n$
for which the coefficient in front of $E_{n-1}$ 
in Eq. (\ref{39}) is not equal to zero. Therefore
\begin{equation}
(N+2j+2)[w+(2N+2j+3)^2]=0
\label{47}
\end{equation}
(compare with Eq. (\ref{33})). As a consequence of
the definition of $n_{max}$, the second multiplier
in this expression is equal to zero if
$N=n_{max}-j$. Therefore the quantity
$N$ is the lesser of $n_{max}-j$ and
$p-2j-2$. Since $j\leq n_{max}$, the only
possibility for $N$ is such that 
\begin{equation}
{\tilde\mu} =2N+2j+3
\label{48}
\end{equation}
Then, as follows
from Eqs. (\ref{45}) and (\ref{46}), 
when $N$ is odd or even, only the term with
$l=[(N+1)/2]$ (where $[(N+1)/2]$ is the integer 
part of $(N+1)/2$) contributes to the sum.

As a consequence
\begin{eqnarray}
&&\Delta_0^N(\lambda)=(-\lambda)^{r(N)}
\prod_{k=1}^{[(N+1)/2]}[\lambda^2+({\tilde\mu}-2j-4k+1)^2]
\label{49}
\end{eqnarray}
where $r(N)=0$ if $N$ is odd and $r(N)=1$ if $N$ is even. 
If $p=3\,\,(mod\,\,4)$, this equation has solutions 
only if $GF(p)$ is extended, and the minimum extension
is $GF({p^2})$. Then the solutions are given by
\begin{equation}
\lambda =\pm i({\tilde\mu}-2j-4k+1)\quad (k=1,2...[(N+1)/2])
\label{50}
\end{equation}
and when $N$ is even there also exists an additional
solution $\lambda=0$. When $N$ is odd (and the dimension
of $V(j)$ is even) the solutions can be represented as
\begin{equation}
\lambda =\pm 2i,\,\pm 6i,...\pm 2iN
\label{51}
\end{equation}
while when $N$ is even, the solutions can be represented as
\begin{equation}
\lambda =0,\, \pm 4i,\,\pm 8i,...\pm 2iN
\label{52}
\end{equation}
Therefore the spectrum is equidistant and the
distance between the neighboring elements is equal to $4i$.
As follows from Eqs. (\ref{48}), all 
the roots are simple. Then, as follows from Eq. (\ref{43}),
the operator $M_{04}$ has a spectral decomposition and
this completes the proof of our main statement (see
Sect. \ref{S1}).

\section{Hermiticity conditions}
\label{S6}

As shown in the preceding sections, for 
physically meaningful modular IRs of the
so(1,4) algebra, the extension of $GF(p)$ to
$GF({p^2})$ guarantees that they are fully 
decomposable. Therefore, as explained in Sect.
\ref{S2}, one can define a scalar product in the 
representation space. By analogy with the
standard theory, we now assume that the
operators $M^{ab}$ are Hermitian with respect
to the chosen scalar product. Therefore, as follows
from Eq. (\ref{22}), the Hermiticity conditions
for the operators $({\bf J}',{\bf J}",R_{kl})$
are as follows
\begin{eqnarray}
(J_-')^*=J_+'\quad (J_-")^*=J_+" \quad 
R_{12}^*=R_{21}\quad R_{11}^*=-R_{22} 
\label{53}
\end{eqnarray}
while the operators $J_3'$ and $J_3"$ should
be Hermitian.

In the spinless case, as follows from Eqs. (\ref{24}), 
(\ref{27}), (\ref{28}) and (\ref{30}),
\begin{equation}
(e^{n+1},e^{n+1})=\frac{n+1}{4(n+2)}[w+(2n+3)^2](e^n,e^n)
\label{54}
\end{equation} 
Therefore, if we assume that $(e^0,e^0)=1$ then
\begin{equation}
(e^n,e^n)=\frac{1}{4^n(n+1)}\prod_{l=1}^n[w+(2l+3)^2]
\label{55}
\end{equation}
It is also easy to show that the elements $e^n$ with
the different values of $n$ are mutually orthogonal.

Our conclusion is as follows. The scalar products of
the basis elements are fully defined by the value of
$(e^0,e^0)$ assuming that the operators $M^{ab}$ are
Hermitian. If $(e^0,e^0)\neq 0$ then all the basis
elements $e^n_{\alpha\beta}$ have a nonzero norm and
are mutually orthogonal. This is in agreement with the
properties of the scalar product in the modular case
(see Sect. \ref{S2}). The quantum numbers 
$(n\alpha\beta)$ characterize the eigenvalues of the
operators $J_3'$, $J_3"$ and ${\bf J}^{'2}$ (note
that in the spinless case the elements 
$e^n_{\alpha\beta}$ are the eigenvectors of the
operators ${\bf J}^{'2}$ and ${\bf J}^{"2}$ with
the same eigenvalues).

Our next goal is to show that the elements 
$e(n,k,l)$ (see Sect. \ref{S4}) have nonzero
norms and are mutually orthogonal. A direct
calculation using Eqs. (\ref{7}), (\ref{11}), 
(\ref{35}), (\ref{36}) and (\ref{55}) gives
\begin{equation}
(e(n,k),e(n,k))=k![\frac{(n-k)!}{n!}]^2
\frac{(2n+1-k)!}{(2n+1-2k)!}(e^n,e^n)
\label{56}
\end{equation}
Then, as follows from the definition of the 
elements $E_n$ and Eq. (\ref{40}),
\begin{equation}
(E_n,E_n)=\frac{1}{4^j(j+1)}\{\prod_{l=1}^j[w+(2l+1)^2]\}
\prod_{l=0}^{n-1}A_{l,l+1}
\label{57}
\end{equation}
In particular, these elements have nonzero norms and
are mutually orthogonal.

Suppose now that $\lambda$ is one of the eigenvalues
given by Eq. (\ref{50}) and $\chi(\lambda)$ is the
eigenvector of $M_{04}$ with this eigenvalue (see
Eq. (\ref{43})). Since $M_{04}$ is Hermitian and
$\lambda$ is imaginary, the only possible value of
$(\chi(\lambda),\chi(\lambda))$ is zero (see Sect.
\ref{S2}). Moreover, since all the eigenvalues are
imaginary when $N$ is odd and there also exists
an additional eigenvalue $\lambda=0$ when $N$ is
even, $(\chi(\lambda_1),\chi(\lambda_2))$ is 
necessarily equal to zero if $\lambda_1+\lambda_2=0$.
Let us show, however, that
$(\chi({\bar\lambda}),\chi(\lambda))\neq 0$ for
imaginary eigenvalues and $(\chi(0),\chi(0))\neq 0$.

As follows from Eqs. (\ref{43}) and (\ref{55}), 
\begin{eqnarray}
&&(\chi({\bar\lambda}),\chi(\lambda))=
\frac{1}{4^j(j+1)}\{\prod_{l=1}^j[w+(2l+1)^2]\}\nonumber\\
&&\{\sum_{n=0}^n \Delta_0^{n-1}(\lambda)^2/
[\prod_{k=0}^{n-1}A_{k,k+1}]\}
\label{58}
\end{eqnarray} 
and one can directly verify that a generalization of
Eq. (\ref{42}) is 
\begin{equation}
\Delta_0^N(\lambda)=\Delta_0^n(\lambda)\Delta_{n+1}^N(\lambda)-
A_{n,n+1}\Delta_0^{n-1}(\lambda)\Delta_{n+2}^N(\lambda)
\label{59}
\end{equation}
since in our case $A_{n+1,n}=1$. Since $\lambda$ is
the eigenvalue, $\Delta_0^N(\lambda)=0$ and one can use Eq.
(\ref{59}) for $k=n,n-1,...0$. As a consequence,
\begin{equation}
\Delta_0^n(\lambda)=\{\prod_{l=0}^nA_{l,l+1}\}
\Delta_{n+2}^N(\lambda)/\Delta_1^N(\lambda)
\label{60}
\end{equation}
and Eq. (\ref{58}) can be rewritten as
\begin{eqnarray}
&&(\chi({\bar\lambda}),\chi(\lambda))=
\frac{1}{4^j(j+1)}\{\prod_{l=1}^j[w+(2l+1)^2]\}\nonumber\\
&&\{\sum_{n=0}^n \Delta_0^{n-1}(\lambda)
\Delta_{n+1}^N(\lambda)\}/\Delta_1^N(\lambda)
\label{61}
\end{eqnarray} 
The sum can be written as
$-d\Delta_0^N(\lambda)/d\lambda$ and therefore, as
follows from Eqs. (\ref{49}) and (\ref{61})
\begin{eqnarray} 
&&(\chi({\bar\lambda}),\chi(\lambda))=
\frac{(-\lambda)^{r(N)}}{4^j(j+1)}\{\prod_{l=1}^j
[w+(2l+1)^2]\}\nonumber\\
&&\{\prod_{\lambda_l\neq \lambda} 
(\lambda - \lambda_l)\}/\Delta_1^N(\lambda)
\label{62}
\end{eqnarray}
where the last product is taken over all the 
eigenvalues excepting $\lambda$. Since all the
eigenvalues are simple (see Sect. \ref{S5})
this product is not equal to zero, and since the
l.h.s. of Eq. (\ref{62}) cannot be singular by
construction, it is not equal to zero.  

We conclude that the basis elements 
$\chi(\lambda_l)$ in $V(j)$ satisfy the following
orthogonality properties. If $\lambda_l$ is imaginary
then ${\bar\lambda}_l$ also is the eigenvalue.
The element $\chi(\lambda_l)$ is orthogonal to
itself and all the other elements $\chi(\lambda_k)$ if
$\lambda_l\neq {\bar\lambda}_k$ while 
$(\chi(\lambda_l),\chi({\bar\lambda}_l))\neq 0$.
When $N$ is even, there also exists the element
$\chi(0)$ which is orthogonal to all the
other elements $\chi(\lambda_l)$ while 
$(\chi(0),\chi(0))\neq 0$.  

\section{Conclusion}
\label{S7}

The main difference between our approach and
the standard one is that we do not postulate
from the beginning that physical states are
elements of a specific linear space. Following
our previous publications, we assume that the
ultimate quantum theory will be based on a
finite field. Then we investigate what is the
minimum extension of the residue field modulo
$p$ such that representations of the symmetry
algebra are fully decomposable. 

When the characteristic of the finite field $p$ is
large, the operators representing physical quantities
act in spaces, the dimensions of which are large.
One can show \cite{lev1,lev3} that
the dimensions of spinless IRs are of order $p^3$.
In the present paper we decompose the representation
space into subspaces $V(j,k)$ the dimensions of which
may be of order $p$. Since finite fields are
not algebraically closed, there is no guarantee
that the characteristic equation of such a
large power will have solutions in the given
finite field.

We believe, it is a very interesting result that
if the symmetry algebra is the modular analog of the
de Sitter algebra so(1,4) then the complex 
extension is already sufficient for ensuring 
spinless IRs to be fully decomposable. This might
also be an explanation of the fact that the present 
quantum theory is based on complex numbers. 

In the literature the operator $M_{04}$ is
usually treated as the de Sitter analog
of the energy operator (see e.g. Refs. \cite{jpa}). 
For this reason one might 
think that the existence of imaginary
eigenvalues of this operator
excludes a possibility that a theory based on a
finite field might be realistic. This problem will
be discussed elsewhere.

In the standard theory the role of
the scalar product is at least threefold: i) to
ensure real eigenvalues of selfadjoint operators;
ii) to ensure spectral decomposition for such
operators; iii) to ensure probabilistic interpretation
in Copenhagen formulation. As it has been already
noted, there are no reasons of why complex eigenvalues
should be excluded, and it has also been shown that
ii) can be valid without assuming the existence of any 
scalar product and Hermiticity requirement. 
The results of Sect. \ref{S6} show that in GFQT 
one can define 
a scalar product in such a way that at least for a 
subset of elements from the representation space,
the probabilistic interpretation is valid. 
However, the problem arises how to interpret the
states which do not have counterparts in the standard
theory (e.g. the states with zero norm). 

{\it Acknowledgements: } The author is grateful to 
H. Doughty, B. Fridman, C. Hayzelden, V.A. Karmanov, G. Mullen, 
and M.A. Olshanetsky for valuable discussions.

\end{document}